\documentclass[seceq]{ptptex}

\usepackage[dvipdfm]{graphicx}
\usepackage{wrapft}
\usepackage{amsmath}
\usepackage{amssymb}

\title{
Coulomb Breakup Reactions\\ in Complex-Scaled Solutions of the
Lippmann-Schwinger Equation%
}

\author{
Yuma \textsc{Kikuchi}$^1$, Takayuki \textsc{Myo}$^2$,
Masaaki \textsc{Takashina}$^3$,\\ Kiyoshi \textsc{Kat\={o}}$^1$
and Kiyomi \textsc{Ikeda}$^4$%
}

\inst{
$^1$Division of Physics, Graduate School of Science,
Hokkaido University,\\Sapporo 060-0810, Japan\\
$^2$General Education, Faculty of Engineering, Osaka Institute of
Technology,\\Osaka 535-8585, Japan\\
$^3$Research Center for Nuclear Physics (RCNP), Osaka University,\\
Ibaraki 567-0047, Japan\\
$^4$The Institute of Physical and Chemical Research (RIKEN),
Wako 351-0198, Japan
}

\abst{
We propose a new method to describe three-body breakups of nuclei, in
which the Lippmann-Schwinger equation is solved combining with the
complex scaling method. The complex-scaled solutions of the
Lippmann-Schwinger equation (CSLS) enables us to treat boundary
conditions of many-body open channels correctly and to describe a
many-body breakup amplitude from the ground state. The Coulomb breakup
cross section from the $^6$He ground state into $^4$He+$n$+$n$
three-body decaying states as a function of the total excitation
energy is calculated by using CSLS, and the result well reproduces the
experimental data. Furthermore, the two-dimensional energy distribution
of the $E1$ transition strength is obtained and an importance of the
$^5$He($3/2^-$) resonance is confirmed. It is shown that CSLS is a
promising method to investigate correlations of subsystems in three-body
breakup reactions of the weakly-bound nuclei.
}

\begin{document}

\maketitle

\section{Introduction}
A neutron halo structure is one of the most interesting topics in
physics of neutron-rich nuclei. In particular, the two-neutron halo
structure observed in the Borromean systems such as $^6$He and $^{11}$Li,
where any binary subsystem does not have a bound state,
has attracted much attention and has been studied by many
authors.~\cite{Ta96,Zh93,Ao06} Theoretically, many works have been
performed to understand a binding mechanism of these nuclei based on
core+$n$+$n$ three-body models, and an importance of two-neutron
correlations has been pointed out.~\cite{Zh93,Ao06,Su91,Be92,Fu94,Ha07}
Recently, it has been discussed how to clarify the internal correlations
of core-$n$ and $n$-$n$ subsystems in the two-neutron halo nuclei from
observables.~\cite{Ch05,Er06}

Coulomb breakup reactions using a high-$Z$ target such as Pb have been
considered as useful tools to investigate the weakly-bound halo nuclei.
For $^6$He, the Coulomb breakup cross sections were measured by
GSI~\cite{Au99} and MSU~\cite{Wa02} groups. For $^{11}$Li, there were
three sets of data measured at MSU~\cite{Iek93}, RIKEN~\cite{Sh95} and
GSI~\cite{Zi97}, and recently, a new measurement at RIKEN~\cite{Na06}
was reported by Nakamura {\it et al}. Through those data, we can obtain
the understanding not only of ground state properties, but also of
breakup reaction mechanisms of halo nuclei. Especially, for two-neutron
halo nuclei, the observed cross section is expected to give some
important information on internal correlations of core-$n$ and $n$-$n$
subsystems. To understand the correlations of subsystems, it is
necessary to investigate the Coulomb breakup reaction based on a
reliable theoretical approach.

In our previous studies, we have successfully described the Coulomb
breakup reactions of $^6$He and $^{11}$Li by using the extended
core+$n$+$n$ three-body model and the complex scaling method
(CSM).~\cite{Myo01,Myo03,My07b} In these analyses, the cross sections
were calculated using the response function method (RFM) combined with
CSM, which is based on the linear response dominated by the $E1$
strength. For $^6$He, the
strength distribution is found to have a peak at around 1 MeV, which is
dominated by the transition into the $^5$He(3/2$^-$)+$n$ two-body
continuum states.~\cite{Myo01} This result indicates that the sequential
breakup process via the $^5$He(3/2$^-$)+$n$ components is important,
and it is shown in Ref.~\citen{Myo01} that the low energy peak
originates from the threshold effect reflecting the halo structure of
the ground state. In the $^{11}$Li breakup case, it was shown that the
non-resonant three-body continuum states of $^9$Li+$n$+$n$ give a
comparable contribution with the sequential process via $^{10}$Li+$n$ in
the cross section.~\cite{My07b} For both cases, our results well
reproduced the observed breakup cross sections with respect to
total excitation energies.

In the previous analysis, correlations of subsystems were investigated
by separating the transition strength into the components of resonant
and non-resonant continuum states. This separation of the strength is
useful when we discuss the effects of resonant and non-resonant
continuum components on the structure of the strength, and further
clarify how much the total strength is exhausted by the individual
strength. While the total strength obtained using RMF reproduces the experimental
observable, however, the separated strength does not correspond to the
observable directly since the experimental data always contains both
contributions of resonant and non-resonant continuum states.
Then, in this study, we consider another approach, which describes the
observables exhibiting the information on internal correlations in
a three-body breakup. To extract internal correlations in a three-body
system from the observable, it is essential to describe the physical
quantities as function of relative energies and momenta in binary
subsystems. In fact, experimentally, the breakup cross section was
reanalyzed as a function of subsystem energies to understand
correlations in two-neutron halo nuclei.~\cite{Ch05}

Theoretically, it is a difficult problem to describe physical quantities
of three-body breakups with composite particles having internal
structures. The standard methods such as the Faddeev, of course, work
well when we handle a simple three-body scattering with point particles.
However, it is difficult to apply them to the composite particle case.
Therefore, an alternative method is needed.

For the two-body case, in Ref.~\citen{Kr07}, it is shown that we can
calculate the exact scattering amplitude by using the formal solution of
the Lippmann-Schwinger equation (LS Eq.) with the complex-scaled Green's
function even if we handle a scattering problem with composite particles.
It is noticed that the Green's function in Ref.~\citen{Kr07} is
constructed by discretized eigenstates of the complex-scaled Hamiltonian,
which are solved in the same manner as bound state cases, and satisfies
correct boundary conditions without any explicit enforcement of boundary
conditions. Furthermore, it is shown that this complex-scaled Green's
function also works to describe the three-body breakups.~\cite{Myo01} It
indicates that we can easily apply the procedure in Ref.~\citen{Kr07} to
three-body cases.

Additionally, the formal solution of the LS Eq. is useful to describe
physical quantities of three-body breakups as functions of relative
energies in subsystems, because it is represented by a solution of an
asymptotic Hamiltonian, namely, a plane wave of a three-body system.

The purpose of this work is to extend the theoretical approach in
Ref.~\citen{Kr07} to three-body systems and propose a new method which
can evaluate scattering observables as functions of energies in
subsystems in three-body breakup reactions. In this paper, we apply this
method to the Coulomb breakup reaction of $^6$He and show that this
method is capable of investigating internal correlations of subsystems
in three-body decaying systems. The reliability of this method is shown
by calculating the Coulomb breakup cross section of $^6$He. Furthermore,
we evaluate the two-dimensional energy distributions of the $E1$
transition strength associated with the subsystems in $^6$He, which is
useful to investigate the internal correlations. In particular, the
importance of the $^5$He(3/2$^-$) resonance in the final states is
confirmed.

This paper is organized as follows. In \S~2, we give an explanation of
our method to describe three-body scattering states as functions of
subsystem energies. In \S~3, we show the obtained results of the Coulomb
breakup reaction of $^6$He, and discuss the reliability of our method
and the correlations of subsystems seen in this reaction. The last
section, \S~4, contains a brief summary.

\section{Complex-scaled solution of Lippmann-Schwinger equation\\
for three-body breakup}
In this section, we explain our new method to describe the three-body
Coulomb breakup reaction in an energy representation of subsystems.
Before describing our method, we give brief explanations of the
$^4$He+$n$+$n$ three-body model of $^6$He and CSM in 2.1 and 2.2,
respectively. In 2.3, we describe the formalism of our method named as
the complex-scaled solutions of the Lippmann-Schwinger equation (CSLS).
\subsection{$^4$He+$n$+$n$ model of $^6$He}
We first explain the $^4$He+$n$+$n$ three-body model of $^6$He briefly.
More detailed explanation is given in Ref.~\citen{Myo01}. In this model,
we describe the $^4$He core as the ($0s$)$^4$ configuration, whose
oscillator length $b_c$ is taken as 1.4 fm to reproduce the charge
radius of $^4$He. In order to analyze the breakup reactions, it is
important to reproduce a threshold energy for each open channel and
scattering properties of every subsystem correctly. Hence, we employ the
orthogonality condition model (OCM)~\cite{Sa68}, in which we can use the
reliable Hamiltonian whose inter-cluster potentials satisfy the
conditions mentioned above.

We solve the following OCM equation for the relative wave function
$\chi^{J^\pi}$ of the $^4$He+$n$+$n$ system;
\begin{equation}
\hat{H} \chi^{J^\pi}(nn) = E \chi^{J^\pi}(nn),
\label{eq:OCM}
\end{equation}
where the Hamiltonian for the relative motion is expressed as
\begin{equation}
\hat{H} = \sum_{i=1}^3 t_i - T_G
+ \sum_{i=1}^2 V_{\alpha n} (\mathbf{r}_i) + V_{nn}
+ V^3_{\alpha nn} + V_\text{PF}.
\label{eq:Ham}
\end{equation}
The operators $t_i$ and $T_G$ describe a kinetic energy of each cluster
and a center-of-mass motion of a three-body system, respectively, and
$\mathbf{r}_i$ ($i=1$ or $2$) represents a relative coordinate between
$^4$He and each valence neutron. The interactions $V_{\alpha n}$ and
$V_{nn}$ are given by the microscopic KKNN potential and the effective
Minnesota potential, respectively. These potentials well reproduce the
scattering data of $^4$He-$n$ and $n$-$n$ systems. In this three-body
model, there is a small deficiency of the binding energy ($\sim$ a few
hundred keV) of $^6$He ground state, which is considered to come from
the $^4$He core polarization effect.~\cite{Myo01} In order to improve
this deficiency of the binding energy, we employ the effective
$\alpha nn$ three-body interaction $V^3_{\alpha nn}$ as
\begin{equation}
V^3_{\alpha nn} = V_3 e^{-\nu (\mathbf{r}_1^2+\mathbf{r}_2^2)},
\end{equation}
where $V_3 = -1.503$ MeV and $\nu = 0.07/b_c^2$ fm$^{-2}$.

The component of the Pauli forbidden state is excluded in the relative
wave function $\chi^{J^\pi} (nn)$ by using the so-called pseudo
potential $V_\text{PF}$. In the case of $^6$He with the $^4$He core, the
Pauli forbidden state $\phi_\text{PF}$ for the valence neutrons is the
occupied $0s$ state of $^4$He. Then, the pseudo potential $V_\text{PF}$
is given as
\begin{equation}
V_\text{PF} =
\sum_{i=1}^2 \lambda | \phi^i_\text{PF} \rangle
\langle \phi^i_\text{PF} |,
\end{equation}
where $\lambda$ is taken as $10^6$ MeV and $i$ is an index for valence
neutrons.

Equation (\ref{eq:OCM}) is solved accurately in a few-body technique. We
here employ the variational hybrid-TV model, in which the relative wave
function of the $^4$He+$n$+$n$ system are expanded on the superposed
basis states of the cluster orbital shell model (COSM; V-basis) and the
extended cluster model (ECM; T-basis)~\cite{Ao95a,Myo01};
\begin{equation}
\chi^{J^\pi} (nn) = \chi^{J^\pi}_V (\boldsymbol{\xi}_V)
+ \chi^{J^\pi}_T (\boldsymbol{\xi}_T),
\end{equation}
where $\chi^{J^\pi}_{V,T} (\boldsymbol{\xi}_{V,T})$ expresses the
relative wave function, and $\boldsymbol{\xi}_V$ and
$\boldsymbol{\xi}_T$ are V- and T-type coordinate sets, respectively.
The radial component of each relative wave function is expanded by
Gaussian basis functions
(Gaussian expansion method, GEM).~\cite{Ao95a,Hi03} This model
successfully describes the observed properties of $^6$He such as the
two-neutron binding energy (0.975 MeV) and the matter radius (2.46 fm)
of the $0^+$ ground state.

\subsection{Complex scaling method}
In CSM~\cite{AC71,Ho83,Mo98}, relative coordinates for a many-body
system are transformed as
\begin{equation}
U(\theta) \mathbf{r} U^{-1}(\theta) = \mathbf{r} e^{i\theta},
\end{equation}
where $U(\theta)$ is a complex scaling operator and $\theta$ is a
scaling angle given in a real number. Applying this transformation to
the Hamiltonian $\hat{H}$, we obtain the complex-scaled Hamiltonian
$\hat{H}^\theta$. For $\hat{H}^\theta$, the corresponding complex-scaled
Schr\"{o}dinger equation is expressed as
\begin{equation}
\hat{H}^\theta \chi^\theta = E\chi^\theta, \ \ \ \
\chi^\theta = e^{(3/2)i\theta\cdot f} \chi(\mathbf{r}e^{i\theta}),
\label{eq:CSS}
\end{equation}
where $\chi^\theta$ is a complex-scaled wave function. The factor
$e^{(3/2)i\theta\cdot f}$ comes from a Jacobian for a volume integral
with $f$ degrees of freedom of a system ($f=2$ for a three-body system).

\begin{figure}[t]
\begin{minipage}{5.3cm}
\centering \includegraphics[width=5.3cm,clip]{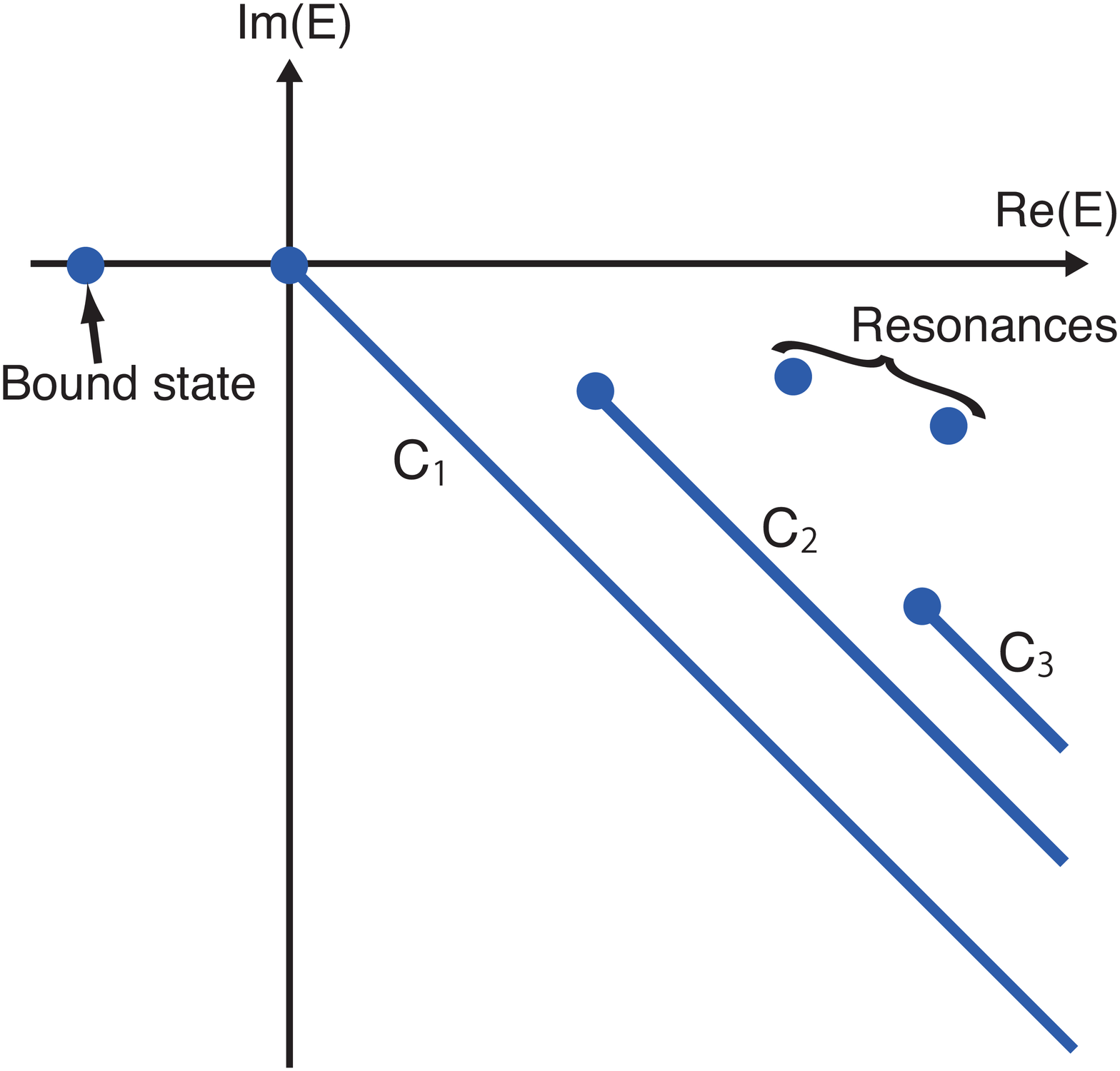}
\end{minipage}
\hfill
\begin{minipage}{8cm}
\centering \includegraphics[width=8cm,clip]{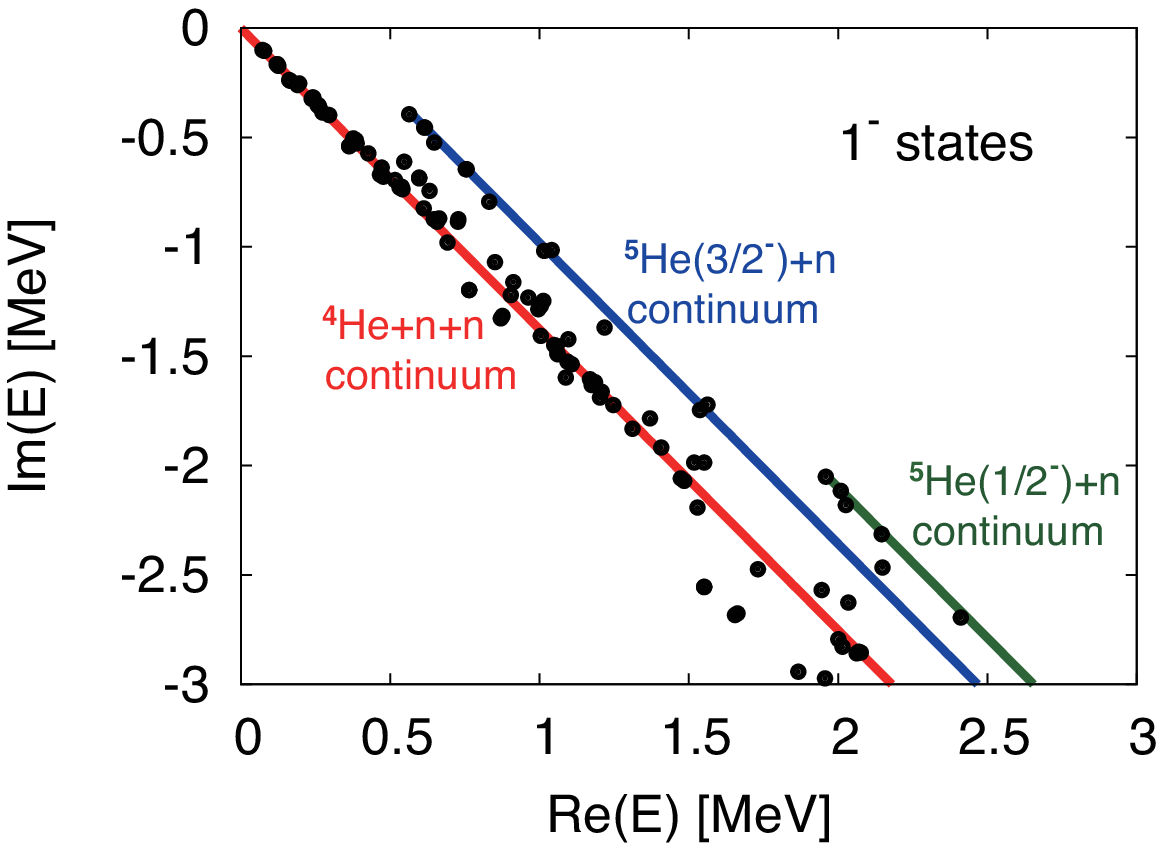}
\end{minipage}
\caption{
Energy spectra of $^6$He in CSM. The left panel is a schematic picture
of spectra of the $^4$He+$n$+$n$ system, where the indices $C_1$, $C_2$
and $C_3$ indicate the $^4$He+$n$+$n$, $^5$He(3/2$^-$)+$n$ and
$^5$He(1/2$^-$)+$n$ continuum spectra, respectively. The right panel is
the $1^-$ spectra calculated within the present $^4$He+$n$+$n$ model.
Three straight lines show the three- and two-body continuum states of
$^4$He+$n$+$n$, $^5$He(3/2$^-$)+$n$ and $^5$He(1/2$^-$)+$n$
corresponding to $C_1$, $C_2$ and $C_3$, respectively. In the $1^-$
spectra calculated with the scaling angle $\theta = 27$ degrees, no
bound and resonant states are obtained.
}
\label{CSE}
\end{figure}

We obtain eigenstates (their biorthogonal states) and energy
eigenvalues of the complex-scaled Hamiltonian $\hat{H}^\theta$ as
$\{\chi^\theta_n\}$ ($\{\tilde{\chi}^\theta_n\}$) and $\{E^\theta_n\}$
with a state index $n$, respectively, by solving the eigenvalue problem
of Eq.~(\ref{eq:CSS}) using a finite number of $L^2$ basis functions.
In CSM, all energy eigenvalues of unbound states are obtained on the
lower half of a complex energy plane, governed by
ABC-theorem~\cite{AC71}, and their imaginary parts represent outgoing
boundary conditions. In ABC-theorem, it is proved that a divergent
behavior at an asymptotic region of resonances is transformed to a
dumping one by CSM. This condition enables us to obtain many-body
resonances by the same calculational way as the bound state case. The
resonances are obtained with the complex energy eigenvalues of
$E^\theta_n = E^r_n-i\Gamma_n/2$, where $E^r_n$ and $\Gamma_n$ are
resonance energies measured from the threshold and decay widths,
respectively, and these energy eigenvalues are independent of the
scaling angle $\theta$. On the contrary, energy eigenvalues of continuum
states are obtained on the branch cuts of the Riemann sheet, which are
rotated down by $2\theta$. This difference of the behaviors between
resonances and continuum states makes the energy eigenvalues of
resonances isolated from continuum states as shown in Fig.~\ref{CSE}.
Actually, CSM has been widely employed as a useful tool to identify the
resonance of two-, three- and four-body systems.~\cite{Ao06,My07c}

Moreover, CSM is also useful to solve a decay problem of a many-body
system. In CSM, the obtained energy eigenvalues of continuum states in
a three-body system are classified into two- and three-body ones by
ABC-theorem. These continuum states are located on the $2\theta$-rotated
branch cuts starting from different thresholds of two- and three-body
decay channels, such as $^5$He+$n$ and $^4$He+$n$+$n$ in the case of
$^6$He. (See Fig.~\ref{CSE}.) The classification of continuum states in
CSM imposes that an outgoing boundary condition for each open channel is
taken into account by imaginary parts of energy eigenvalues. Using this
classification of continuum states, we can describe three-body
scattering states without any explicit enforcement of boundary
conditions.

\subsection{Complex-scaled solutions of the Lippmann-Schwinger equation}
We explain a new method of CSLS to describe three-body scattering
states, which is capable of calculating physical quantities as functions
of subsystem energies in a three-body system.

The formal solution of the Lippmann-Schwinger equation can be described
as
\begin{equation}
\Psi^{(\pm)} = \Phi_0 + \lim_{\varepsilon \to 0}
\frac{1}{E-\hat{H}\pm i\varepsilon} \hat{V} \Phi_0,
\label{eq:LS}
\end{equation}
where $\Phi_0$ is a solution of an asymptotic Hamiltonian $\hat{H}_0$.
The total Hamiltonian $\hat{H}$ is given in Eq.~(\ref{eq:Ham}) for
$^6$He and the interaction $\hat{V}$ is given by subtracting $\hat{H}_0$
from $\hat{H}$.
The boundary condition of the scattering state is represented by
$\pm\varepsilon$.

When we consider the scattering states of the Borromean system, the
asymptotic Hamiltonian $\hat{H}_0$ is equivalent to a free Hamiltonian
in a three-body system since all the scattering states are described by
three-body scattering states and any binary subsystem does not have a
bound state.
Then, we obtain the following relations:
\begin{align}
\hat{H}_0 &= \sum_{i=1}^3 t_i - T_G  \\
\Phi_0 &= \frac{1}{(2 \pi^3)}\exp(i \mathbf{k} \cdot \mathbf{r})
\exp(i \mathbf{K}\cdot \mathbf{R}),
\end{align}
where $\mathbf{k}$, $\mathbf{K}$ and $\mathbf{r}$, $\mathbf{R}$ are
relative momenta and relative coordinates in a three-body Jacobi
coordinate system, respectively. We denote $\Phi_0$ by
$\langle \mathbf{k}, \mathbf{K} |$ and
$| \mathbf{k}, \mathbf{K} \rangle$ in the bra- and ket-state
representation, respectively, to describe the momenta in asymptotic
region, $\mathbf{k}$ and $\mathbf{K}$, explicitly.

The formal solution
in Eq.(\ref{eq:LS}) for an outgoing three-body scattering state can be
rewritten in the ket-state representation as
\begin{equation}
| \Psi^{(+)} (\mathbf{k},\mathbf{K}) \rangle
= | \mathbf{k}, \mathbf{K} \rangle
+ \lim_{\varepsilon\to 0}\frac{1}{E-\hat{H}+i\varepsilon} \hat{V}
| \mathbf{k}, \mathbf{K} \rangle,
\label{eq:LSF}
\end{equation}
where the interaction $\hat{V}$ is
\begin{equation}
\hat{V} = \hat{H} - \hat{H}_0
= \sum_{i=1}^2 V_{\alpha n} (\mathbf{r}_i) + V_{nn} +V^3_{\alpha nn}
+ V_\text{PF}.
\end{equation}

In the present work, we use the complex-scaled Green's function
$\mathcal{G}^\theta(E; \boldsymbol{\xi}, \boldsymbol{\xi}^\prime)$.
The complex-scaled Green's function is related to the non-scaled
Green's function
$\mathcal{G}(E; \boldsymbol{\xi},\boldsymbol{\xi}^\prime)$ as
\begin{equation}
\lim_{\varepsilon\to 0} \frac{1}{E-\hat{H}+i\varepsilon}
= \mathcal{G}(E; \boldsymbol{\xi},\boldsymbol{\xi}^\prime)
= U^{-1}(\theta)
\mathcal{G}^\theta (E; \boldsymbol{\xi}, \boldsymbol{\xi}^\prime)
U(\theta),
\label{eq:GF}
\end{equation}
where the complex-scaled Green's function
$\mathcal{G}^\theta (E; \boldsymbol{\xi}, \boldsymbol{\xi}^\prime)$ is
defined as
\begin{equation}
\mathcal{G}^\theta(E; \boldsymbol{\xi}, \boldsymbol{\xi}^\prime)
=\left\langle \boldsymbol{\xi} \left|
\frac{1}{E-\hat{H}^\theta}
\right| \boldsymbol{\xi}^\prime \right\rangle
= \sum_n \hspace{-0.46cm}\int
\frac{\chi^\theta_n(\boldsymbol{\xi})
\tilde{\chi}^\theta_n(\boldsymbol{\xi}^\prime)}
{E-E^\theta_n}.
\label{eq:CSGF}
\end{equation}
In derivation of the right hand side of Eq.~(\ref{eq:CSGF}), we use the
extended completeness relation (ECR), whose detailed explanation is
given in Ref.~\citen{Myo98} and skipped here. Using this Green's
function constructed by the complex-scaled wave functions of bound,
resonant and non-resonant continuum states, we can take into account
boundary conditions for all open channels of a three-body system in the
form of complex energy eigenvalues $E^\theta_n$. Then, we can omit the
operation of $\varepsilon\to 0$ in the derivation of the complex-scaled
Green's function in Eq.~(\ref{eq:CSGF}).
It should be noted that the Green's function in Eq.~(\ref{eq:CSGF}) is a
continuous function with respect to the total energy $E$ while we use
discretized energy eigenvalues $E^\theta_n$ of the complex-scaled
Hamiltonian $\hat{H}^\theta$. In this calculation, we use 15 Gaussian
bases for each relative coordinate, the range of which is taken up to
about 20 fm, and the scaling angle $\theta$ is taken as 18 degrees to
construct the Green's function.

Combined with the complex-scaled Green's function in Eq.~(\ref{eq:CSGF}),
the formal solution in Eq.~(\ref{eq:LSF}) is rewritten as
\begin{equation}
| \Psi^{(+)} (\mathbf{k}, \mathbf{K}) \rangle
= | \mathbf{k}, \mathbf{K} \rangle
+ \sum_n \hspace{-0.46cm}\int U^{-1}(\theta)
| \chi^\theta_n \rangle \frac{1}{E-E^\theta_n}
\langle \tilde{\chi}^\theta_n |
U(\theta) \hat{V} | \mathbf{k}, \mathbf{K} \rangle.
\label{eq:CSLS}
\end{equation}
It is not necessary to apply the complex
scaling to the first term of the solution of an asymptotic Hamiltonian.
The complex scaling operator $U(\theta)$ in Eq.~(\ref{eq:CSLS}) is
processed in the calculation of the matrix elements and does not
operate on the complex-scaled eigenstates $\chi^\theta_n$.

Similarly, let us consider the formal solution for an incoming
scattering state.
To describe the incoming scattering state, we here consider the
bra-state of $\Psi^{(-)}$ with momenta $(\mathbf{k},\mathbf{K})$, which
is given as
\begin{equation}
\begin{split}
\langle \Psi^{(-)}(\mathbf{k},\mathbf{K}) |
=& \langle \mathbf{k}, \mathbf{K} |
\left(1+\lim_{\varepsilon\to 0}\frac{1}{E-\hat{H}-i\varepsilon}
    \hat{V}\right)^\dagger
\\=& \langle \mathbf{k}, \mathbf{K} |
+\langle \mathbf{k}, \mathbf{K} | \hat{V} \lim_{\varepsilon\to 0}
\frac{1}{E-\hat{H}+i\varepsilon}.
\label{eq:iwf}
\end{split}
\end{equation}
In the derivation of the second line, we assume the hermiticity of
$\hat{H}$ and $\hat{V}$.
The Green's function in Eq.~(\ref{eq:iwf}) is equal to that of
Eq.~(\ref{eq:GF}), and hence we replce this Green's function in to the
complex-scaled Green's function. Using Eqs.~(\ref{eq:GF}) and
(\ref{eq:CSGF}), we obtain the bra-state of the incoming scattering
state as
\begin{equation}
\langle \Psi^{(-)} (\mathbf{k},\mathbf{K}) |
= \langle \mathbf{k}, \mathbf{K} |
+ \sum_n\hspace{-0.46cm}\int\hspace{0.1cm}
\langle \mathbf{k}, \mathbf{K} | \hat{V} U^{-1} (\theta)
| \chi^\theta_n \rangle \frac{1}{E-E^\theta_n}
\langle \tilde{\chi}^\theta_n | U(\theta).
\label{eq:CSLSi}
\end{equation}
Hereafter, we refer Eqs.~(\ref{eq:CSLS}), (\ref{eq:CSLSi}) and their
conjugate states as the
complex-scaled solutions of Lippmann-Schwinger equation (CSLS).

Additionally, we switch off the pseudo potential $V_\text{PF}$ in the
calculation of Eqs.~(\ref{eq:CSLS}) and (\ref{eq:CSLSi}) to avoid an
instability of numerical
results, which comes from the large value of $\lambda = 10^6$ MeV, while
the wave functions $\chi^\theta_n$ are solved with the pseudo potential.
By switching off the pseudo potential, the antisymmetrization in the
scattering state is approximately ignored, but it is not a serious
problem, since the antisymmetrization in the intermediate states are
considered when we solve the eigenstates $\chi^\theta_n$. In fact,
it will be shown in the next section that the obtained result in CSLS
shows a reasonable agreement with the previous result in
Ref.~\citen{Myo01} and the calculated breakup cross section well
reproduces the trend of the experimental data.


\subsection{E1 transition in complex-scaled solutions of the
Lippmann-Schwinger equation}
To calculate the $E1$ transition strength and the Coulomb breakup cross
section, we start with the two-dimensional momentum distribution of the
$E\lambda$ transition strength given as
\begin{equation}
\frac{d^6B(E\lambda)}{d\mathbf{k} d\mathbf{K}}
= \frac{1}{2J_\text{g.s.}+1}
\left| \langle \Psi^{(-)} (\mathbf{k}, \mathbf{K}) ||
\hat{O} (E\lambda) || \Phi_\text{g.s.} \rangle \right|^2,
\label{eq:MD}
\end{equation}
where $\Phi_\text{g.s.}$ is a ground state wave function and
$\hat{O} (E\lambda)$ is a transition operator with a rank $\lambda$.
$J_\text{g.s.}$ is a total spin of the ground state.

Using Eq.~(\ref{eq:MD}), we derive an $E\lambda$ transition strength
with respect to the total energy $E$ of a system as follows;
\begin{equation}
\frac{dB(E\lambda)}{dE}
= \iint d\mathbf{k} d\mathbf{K}
\frac{d^6 B(E\lambda)}{d\mathbf{k} d\mathbf{K}}
\delta \left( E-\frac{\hbar^2 k^2}{2\mu}-\frac{\hbar^2 K^2}{2M} \right),
\label{eq:ED}
\end{equation}
where $\mu$ and $M$ are reduced masses of subsystems corresponding to
the two momenta $\mathbf{k}$ and $\mathbf{K}$, respectively. Similarly,
we obtained the two-dimensional energy distribution as
\begin{equation}
\frac{d^2 B(E\lambda)}{d\varepsilon_1 d\varepsilon_2}
= \iint d\mathbf{k} d\mathbf{K}
\frac{d^6 B(E\lambda)}{d\mathbf{k} d\mathbf{K}}
\delta \left( \varepsilon_1-\frac{\hbar^2k^2}{2\mu} \right)
\delta \left( \varepsilon_2-\frac{\hbar^2K^2}{2M} \right),
\label{eq:TD}
\end{equation}
where $\varepsilon_1$ and $\varepsilon_2$ are subsystem energies in a
three-body system.

In the next section, we shall show the total energy and
two-dimensional energy distributions of the $E1$ transition strength,
and discuss the correlations of subsystems in the Coulomb breakup of
$^6$He.

\section{Energy distributions of $E1$ transition strength for $^6$He}
We demonstrate that CSLS is useful to investigate internal correlations
of subsystems in the final states.

\begin{figure}[t]
\centering{\includegraphics[width=10cm,clip]{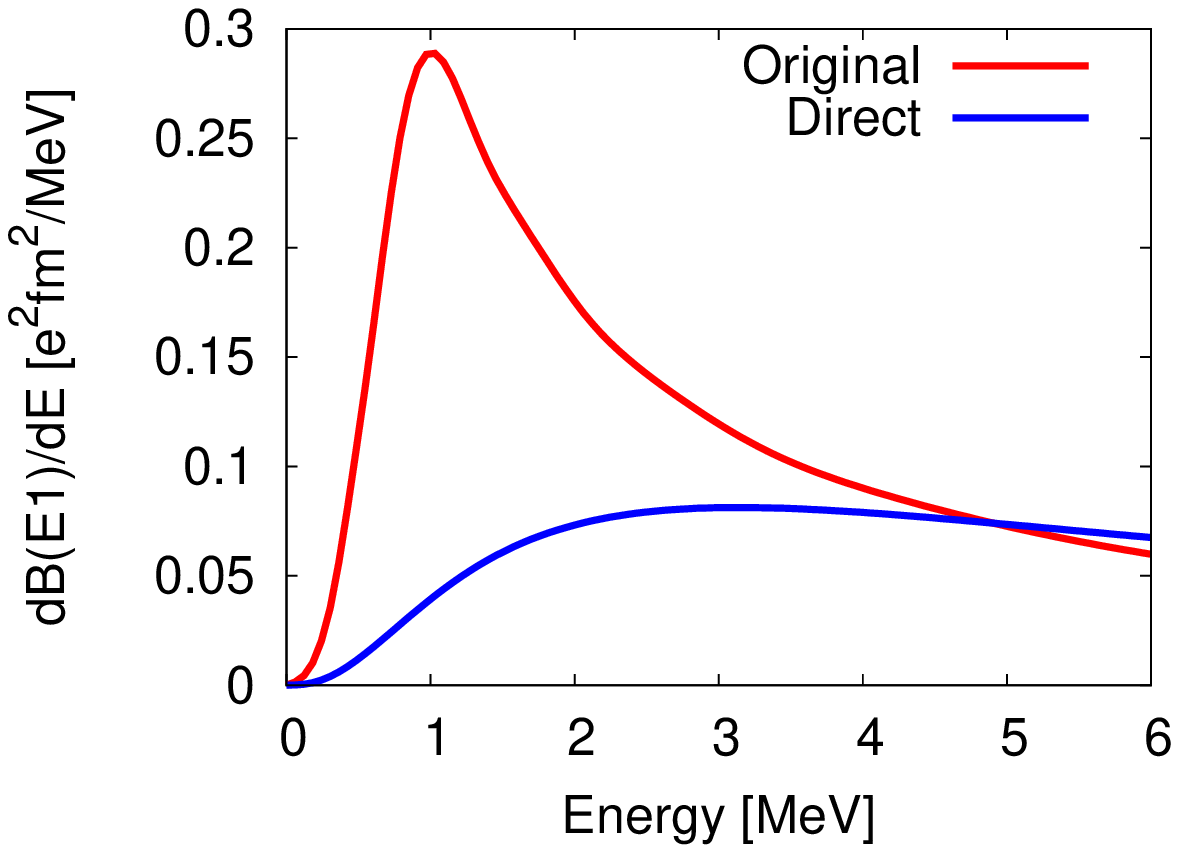}}
\caption{(Color online)
Total energy distribution of the $E1$ transition strength of $^6$He. The
red and blue curves show the result including FSI and the one of 
the direct breakup, respectively.}
\label{FSI_dis}
\centering{\includegraphics[width=10cm,clip]{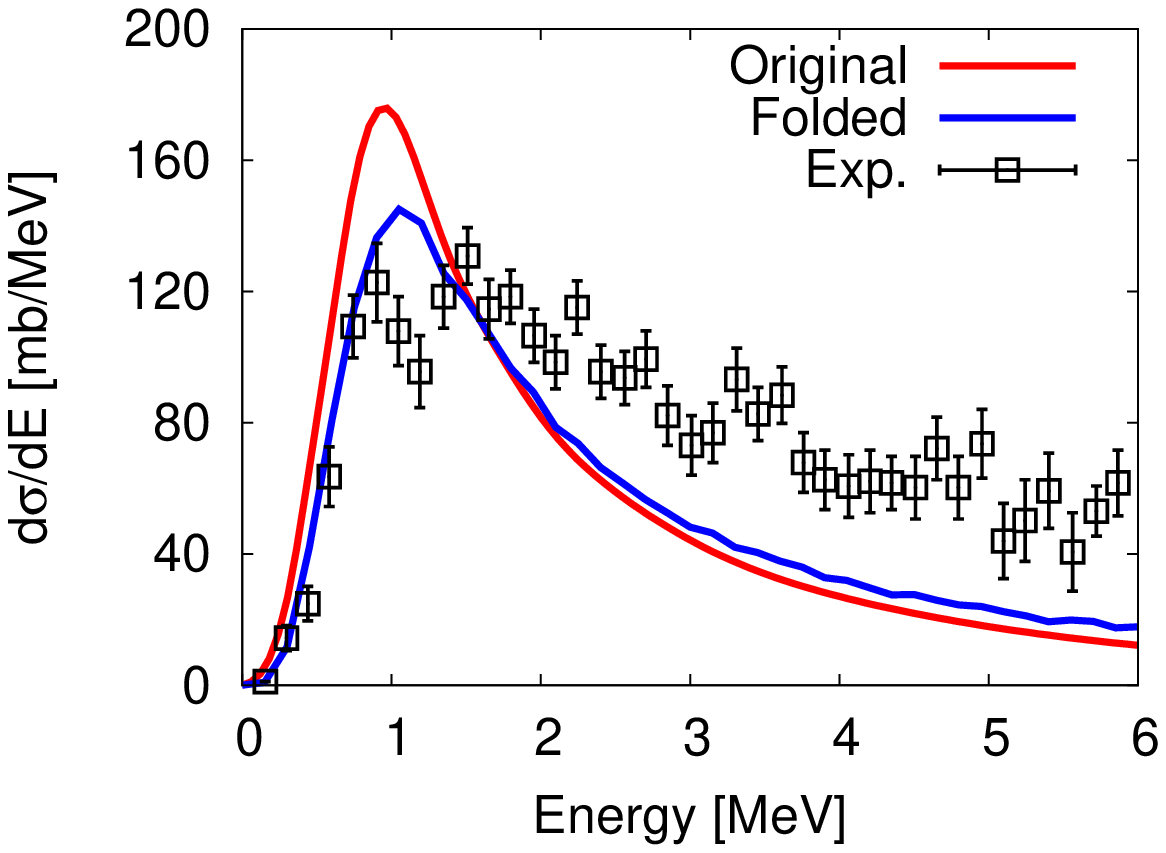}}
\caption{(Color online)
The obtained Coulomb breakup cross section measured from the
$^4$He+$n$+$n$ threshold energy. The red and blue lines show the
original result in CSLS and the folded one with the experimental
resolution, respectively. The observed cross section (open square) is
taken from Ref.~\citen{Au99}.}
\label{E_dis}
\end{figure}
Before discussing the correlations of subsystems in the final states, we
first calculate the total energy distributions of the $E1$ transition
strength from the ground state of $^6$He into $^4$He+$n$+$n$ three-body
scattering states to show the importance of the final state interaction
(FSI). Using Eq.~(\ref{eq:ED}), we obtain the result of the total
energy distribution measured from the $^4$He+$n$+$n$ threshold energy
as shown in Fig.~\ref{FSI_dis}. From this original result, it is
confirmed that the strength possesses a sharp peak at around 1 MeV. In
order to recognize FSI, we also calculate the strength without FSI.
When we switch off FSI, scattering state of $^6$He can be described
only by first term of the right hand side of Eq.~(\ref{eq:CSLS}) since
the interaction $\hat{V}$ is zero. Then, the strength without FSI,
which is equivalent to the one of the transition from the ground state
into non-interacting three-body continuum states, is calculated by
taking the first term of Eq.~(\ref{eq:CSLS}). This transition strength of
the direct breakup is also shown in Fig.~\ref{FSI_dis}. It is found that
the direct breakup one has a small strength with a broad bump at around
3 MeV. This large difference between the original and the direct breakup
strengths indicates the importance of FSI to explain the low energy
enhancement in the $E1$ transition strength of $^6$He. The result of
Fig.~\ref{FSI_dis} is consistent with the previous result in
Ref.~\citen{Myo01}.

Here, we also show the reliability of our calculation by comparing the
obtained result using CSLS and the experimental data. We drive the
breakup cross section of $^6$He using the obtained $E1$ transition
strength and the equivalent photon method. The experimental data are
taken from Ref.~\citen{Au99}. In Fig.~\ref{E_dis}, two strengths are
shown; One is the original result in CSLS and another is the folded one
by the experimental resolution~\cite{Au99}. The obtained cross section
in CSLS has a peak at around 1 MeV and well reproduces the observed
trend. This good agreement of the cross section implies the reliability
of CSLS to describe the three-body scattering states since the validity
of the three-body model has been already shown in Ref.~\citen{Myo01}.

\begin{figure}[t]
\centering{
\includegraphics[width=8cm,clip]{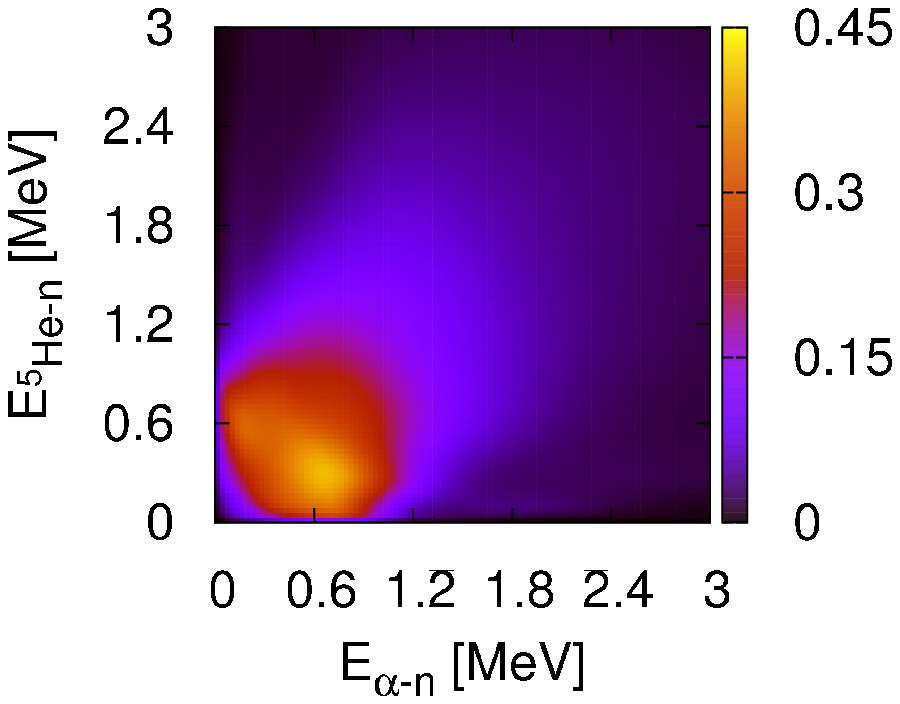}}
\caption{(Color online)
Twi-dimensional energy distribution of the $E1$ transition strength
corresponding to the $^4$He-$n$ and $^5$He-$n$ subsystems.
}
\label{5He_dis}
\centering{
\includegraphics[width=8cm,clip]{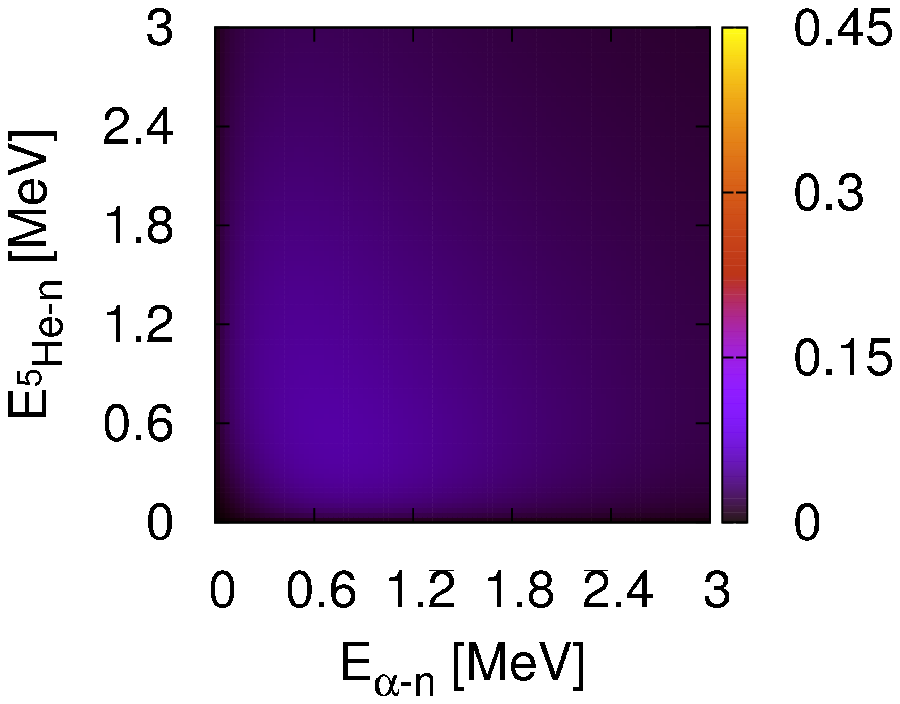}}
\caption{(color online)
Two-dimensional energy distribution of the direct breakup as same as
Fig.~\ref{5He_dis}.
}
\label{free_dis}
\end{figure}
Next, we investigate correlations of subsystems in the final states.
Using Eq.~(\ref{eq:TD}), we evaluate the two-dimensional energy
distribution of the $E1$ transition strength of $^6$He, associated with
the $^5$He subsystem. The result is shown in Fig.~\ref{5He_dis}, where
$\varepsilon_1$ and $\varepsilon_2$ in Eq.~(\ref{eq:TD}) are the
relative energies of $^4$He-$n$ ($E_{\alpha\text{-}n}$) and $^5$He-$n$
($E_{^5\text{He-}n}$) systems, respectively. It is clearly seen that
the strength is concentrated on around $E_{\alpha\text{-}n} \sim 0.7$
MeV, which agrees with the $^5$He(3/2$^-$) resonance energy. Hence, the
importance of the $^5$He(3/2$^-$) resonance in the Coulomb breakup
reaction of $^6$He is directly shown in the physical observables using
CSLS, which is consistent with the previous analysis using
RFM.~\cite{Myo01} We also show the two-dimensional energy distribution
of the direct breakup strength in Fig.~\ref{free_dis} to clarify the FSI
in the two-dimensional energy distributions. From Fig.~\ref{free_dis},
we find that no clear peak structure appears without FSI. This result
indicates that the correlations in the two-dimensional energy
distribution mainly comes from the FSI and the sequential decay via
the $^5$He(3/2$^-$)+$n$ channel is important in the Coulomb breakup
reaction of $^6$He.

\section{Summary}
In the present study, we developed a new approach called the
complex-scaled solutions of Lippmann-Schwinger equation (CSLS), which
enables us to describe a scattering state for three-body breakups of the
Borromean system and to calculate the observables with respect to not
only the total energy but also the subsystem energies. Using CSLS, we
reproduced the observed Coulomb breakup cross section nicely, and
confirmed that the $^5$He(3/2$^-$) resonance plays an important role in
the Coulomb breakup reaction. This CSLS approach makes us to extract the
correlations of subsystems from the observables and is useful to study
the properties of the weakly-bound nuclei not only for the ground state,
but also for the scattering states. The detailed analysis on the
subsystems correlations such as $^4$He-$n$ and $n$-$n$ systems in the
Coulomb breakup reaction of $^6$He is forthcoming. It is also
interesting to perform the analysis of the two-neutron halo nuclei
$^{11}$Li using this method.

\section*{Acknowledgements}
We thank the Yukawa Institute for Theoretical Physics at Kyoto
University for discussions during the YITP workshop YITP-W-06-17 on
Nuclear Cluster Physics. One of the authors (Y. Kikuchi) would like to
thank members of the nuclear theory group at Hokkaido University and
Prof. A. Ohnishi at YITP. This work was supported by Grant-in-Aid for
JSPS Fellow (No. 204495).

%

%

\end{document}